\documentclass{amia}
\usepackage{graphicx}
\usepackage[labelfont=bf]{caption}
\usepackage[superscript,nomove]{cite}
\usepackage{color}
\usepackage{amssymb}
\usepackage{booktabs}
\usepackage{multirow}
\PassOptionsToPackage{hyphens}{url}\usepackage{hyperref}
\begin{document}

\title{Deep Representation for Patient Visits from Electronic Health Records}

\author{Jean-Baptiste Escudi\'e, MD, MSc$^{1}$, Alaa Saade, PhD$^{2}$,\\ Alice Coucke, PhD$^{2}$, Marc Lelarge, PhD$^{3}$}

\institutes{$^1$APHP, Paris, France; $^2$SNIPS, Paris, France ; $^3$INRIA, Paris, France\\
}

\maketitle

\noindent{\bf Abstract}

We show how to learn low-dimensional representations (embeddings) of patient visits from the corresponding electronic health record (EHR) where International Classification of Diseases (ICD) diagnosis codes are removed. We expect that these embeddings will be useful for the construction of predictive statistical models anticipated to drive personalized medicine and improve healthcare quality. These embeddings are learned using a deep neural network trained to predict ICD diagnosis categories. We show that our embeddings capture relevant clinical informations and can be used directly as input to standard machine learning algorithms like multi-output classifiers for ICD code prediction. We also show that important medical informations correspond to particular directions in our embedding space. 

\section{Introduction}

Over the past 10 years, hospital adoption of electronic health record (EHR) system has risen to an unprecedented level. According to the latest report from the Office of the National Coordinator for Health Information Technology, nearly 84\% of hospitals have adopted at least a basic EHR system in the USA\cite{onchit}.
EHR systems store data associated with each patient encountered and are primarily designed for improving healthcare efficiency from an operational standpoint. This paper explores secondary use of EHRs \cite{botsis} in a deep learning framework and builds a representation of the raw data which will be the first required step for any further deep learning analysis.

The representation of raw data is a fundamental issue spanning all types of machine learning frameworks.
Traditionally, input features to a machine learning algorithm are hand-crafted from raw data, relying on practitioner expertise and domain knowledge to determine explicit patterns of interest. The engineering process of creating and selecting appropriate features is laborious and time consuming. In contrast, deep learning techniques learn optimal features directly from the raw data itself without human guidance, allowing for the automatic discovery of latent data structures.

In this paper, we develop a supervised deep learning algorithm to learn low-dimensional representations (also called embeddings) of patient visits. EHR data is challenging to represent and model due to its high dimensionality, noise and sparseness\cite{weiskopf}. A common first step for mining EHR consists for a domain expert to designate the patterns to look for (i.e. the learning task) and to specify the appropriate clinical variables (i.e. the input features). Although appropriate in some cases, this methodology scales poorly, does not generalize well and misses opportunities to discover new patterns.
To address these shortcomings, we design an auxilary machine learning task which consists in predicting the ICD codes of each visit. We build a deep neural network (NN) architecture encoding the raw data into an embedding and train this NN to predict the presence or absence of ICD codes. As a result, we obtain an encoder mapping raw data to a dense vector of $\mathbb{R}^n$ where $n$ is a free parameter (fixed to $448$ in the rest of the paper, see below).

The International Classification of Dieseases (ICD) is a health care classification system maintained by the World Health Organization, which provides a hierarchy of diagnostic codes of diseases, disorders, injuries, signs, symptoms, etc. Given a clinical context, reported in the EHR in the form of free text and structured data, appropriate ICD codes are attributed manually by the physician or other healthcare professionals following the coding guidelines. In our work, we use ICD coding as a form of supervision (i.e. labeling of our dataset) in order to allow our algorithm to learn a valuable representation of the raw data. This approach is similar to recent use of image inpainting in the computer vision community to learn context\cite{pathak}. Our algorithm is asked to produce low-dimensional embeddings of the raw data in order to predict some information withheld intentionally.

In this paper, we build a generalist embedding of patient visits. Although we use the task of ICD codes prediction to do so, this is not our main goal. Indeed, we demonstrate that our representation is a first step towards developing quantitative models for patients than can be used to predict health status, as well as to help preventing diseases or disabilities. In order to test the validity of our representation, we show that it contains a compressed version of the medical informations of the EHR.

\section*{Contributions}

We present a deep learning approach to build a low-dimensional representation for patient visits based on the raw EHR data. Specifically, our NN takes as inputs both clinical free-text notes and strucutred and semi-structured data present in the EHR. Using the MIMIC-III dataset\cite{mimic}, we demonstrate the medical pertinence of the representations found by our algorithm.

\section{Related work}

The use of deep learning on EHR increased rapidly after adoption of EHR systems\cite{onchit} and development of deep learning methods\cite{lecun}. In a well-known work\cite{deeppatient}, authors develop a framework called `deep patient' to represent patients by a set of general features, which are inferred automatically from a large-scale EHR database through a stack of denoising auto-encoders. To prove the effectiveness of the proposed representation, deep patient is used to predict future diseases. As opposed to our approach, the representation built for deep patient\cite{deeppatient} is done in an unsupervised manner. In particular, ICD codes are given as inputs of the algorithm, so that deep patient cannot be used for automated ICD coding. Subsequent work extends this approach by modeling the temporal sequence of events that occured in a patient's record with convolutional networks\cite{lipton}.

Very recently, based on data from two academic hospitals with a general patient population, authors\cite{ehrgoogle} demonstrate the effectiveness of using deep learning models in a wide variety of predictive problems and settings. Three deep learning NN are used. In contrast to our approach, these three NN need to be trained for each separate task (such as prediciting in-hospital mortality, 30-day unplanned readmission, etc.) which allows to get very good performances at the expense of a high computational cost. Moreover, no generalist embedding like ours is learnt following this methodology.

Although this is not our main aim here, there are works towards automated ICD coding. A recent paper\cite{icdpetuum} formulates the coding task as a general multi-label classification problem on diagnosis descriptions and uses a recurrent NN on the free-text data to assign codes.

Given rapid developments in this field, we point readers to a recent review\cite{review}.

\section{Methods}

In this section, we describe our strategy to prepare the dataset and build features as well as labels for each patient stay. We then describe in detail our deep learning approach to assign a vector embedding for each stay.

\subsection{Dataset and preprocessing}
We perform the study on the publicly available MIMIC-III dataset\cite{mimic}, which contains de-identified and comprehensive electronic medical records of $58,976$ patient visits in the Beth Israel Deaconess Medical Center from 2001 to 2012.

To each stay, we associate two types of features and a vector of labels
\begin{itemize}
    \item \emph{Text features}: the MIMIC-III dataset contains a large corpus of medical records made of observations and notes written by care providers during the patient's stay. We use a vector of integers to represent the medical records associated with each stay.
     Specifically, we define a vocabulary of $138\,942$ words with at least $5$ occurrences in the medical records of the whole MIMIC-III dataset. We represent each medical record as a sequence of integers representing the index of each word in this vocabulary, and concatenate these sequences to obtain one such vector of integers for each stay. We truncate these vectors to a maximum length of $16\,618$ words corresponding to the $90$th percentile of medical reports lengths over all stays.
    \item \emph{Structured features}: we also make use of the large amount of numerical information available for each stay, such as the type and quantity of medications given to the patient, with associated severity and mortality scores, time-dependent vital signs, fluid balance, laboratory results, etc. As a first coarse-grained approach, we associate to each stay a vector of structured features by concatenating all these features and summing them over the time of the stay for those that are time-dependent. This yields a vector of approximately $44\,000$ real-valued features for each stay. We note that we exclude from the features the ICD codes, as these will serve as our targets for the learning task (see below).
    \item \emph{Labels}: each discharge contains a set of ICD codes assigned by medical care providers and used for billing purposes. We use these codes as a summary of the medical condition of the patient that can be predicted from the two aforementioned types of features. ICD codes have a hierarchical structure that allow for a variable level of precision in the description of the medical condition of the patient. In the following, we restrict our study to the lowest resolution level, consisting of $19$ so-called `chapters'. Our label vector for each stay is therefore a $19$-dimensional binary vector indicating whether each code has been assigned to the stay of not.
\end{itemize}

As explained in the introduction, EHR data is very sparse. This is illustrated in Table \ref{tab_spars}, where we have grouped together features by type, such as \emph{demographics}, which contains categorical data such as gender, age and ethnicity information, among a total of $54$ different features.
Importantly, these features have a very variable prevalence in the dataset, the most represented value being present in $97.2\%$ of the stays while half of the values are present in only $0.1\%$ of the stays, as quantified by the median frequency of each value shown in the table. Table \ref{tab_spars} shows that other groups of features are similarly sparse.


 There is a high number of lab tests and drugs, microbiology and drug prescriptions which represents most of the features, respectively $19\,457$ and $13\,715$ features. Each such feature is the number of lab tests or prescriptions for each category during the stay. In particular, we ignore when these tests or prescriptions have been done and in this paper, we only count them.

\begin{table}[h!]
\begin{center}
\begin{tabular}{ |c|c|c|c| } 
 \hline
  &  Number of features& Median frequency& Most frequent \\ 
 \hline
 demographics &  54&0.1\%&97.2\% \\ 
 \hline
 administrative &  90&3.6\%&97.2\% \\ 
 \hline
 microbiology &  19547&0.0\%&82.2\% \\ 
 \hline
 input-output &  1871&0.0\%&45.2\% \\ 
 \hline
 prescriptions &  13715&0.0\%&83.2\% \\ 
 \hline
 icd9 procedures &  2683&0.0\%&57.0\% \\ 
 \hline
\end{tabular}
\end{center}
\caption{Sparsity of EHR data}
  \label{tab_spars}
\end{table}

In order to reduce the dimensionality of the structured features, it has proved useful to perform some feature selection. Of the initial $44\,000$ structured features, we discard all but the $8\,000$ best predictors of the ICD codes of the stay according to a univariate $\chi^2$ test. 

\subsection{Model design}

We learn a representation of the patients visits in a supervised way. More precisely, we use a hybrid architecture made of a convolutional neural network on the text features, and a multi-layer perceptron on the structured data, both trained jointly to predict the ICD labels associated with each stay. This yields a multi-label classification task in which we predict, for each code, whether it was assigned to the stay of not. Finally, we extract an embedding for each stay by concatenating the output of the last hidden layer of the two subnetworks.

\begin{figure}[h!]
\centering
\includegraphics[scale=0.4]{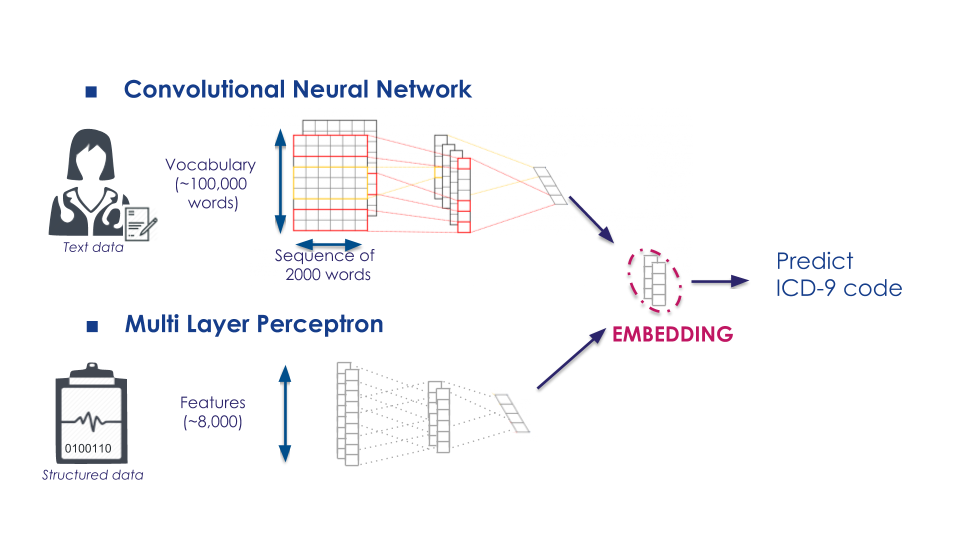}
\caption{General description of the Neural Network architecture}
\label{model}
\end{figure}

The resulting architecture is pictured on Figure \ref{model}, and the following subsections describe the two subparts of the network in more details. Importantly, we note that this multi-label classification task is not the main object of our work. In fact, we use the ICD codes as mere proxies to design an efficient supervised training strategy to learn our embeddings. In Section \ref{results_section}, we shall validate our approach both by comparing the accuracy of our classification algorithm against baseline models, and independently illustrate the medical relevance of our embeddings.

\subsubsection{Multi-layer perceptron (MLP)}

Our model for the structured features is a standard $2$-layers perceptron, with a hidden layer of size $256$. We use rectified linear units (ReLU) \cite{relu_study} as our activation functions on the input and hidden layers. Dropout \cite{dropout} is applied after the hidden layer to reduce overfitting. The output of the structured features model is a $19$-dimensional vector where each entry corresponds to a score for the corresponding ICD chapter. The probability of this chapter being assigned to the sample is given by the sigmoid of this score.

\subsubsection{Convolutional neural network (CNN)}

Our convolutional architecture for text features is inspired by recent developments in text classification\cite{cnn_text_classif}. Specifically, we start by embedding each word in the input sentence in a $50$-dimensional space using a dense lookup table. We then apply a one-dimensional convolutional layer with three types of filters of size $(3, 4, 5)\times 50 \times 64$, where $3,4$ or $5$ are the sizes of the receptive fields, $50$ is the word embedding dimension and $64$ is the number of channels. We then use max-pooling over the time dimension to obtain a vector representation of the medical report of size $3\times 64 = 192$, and add a dropout layer. Finally, we obtain a $19$-dimensional vector of probabilities of each ICD code by applying a fully connected layer followed by a sigmoid non-linearity.

\subsubsection{Representation learning}

The output probability vector of the hybrid network is obtained by summing component-wise the $19$-dimensional probability vectors of the multi-layer perceptron and the convolutional neural network. We consider a multi-label classification setting and define our loss to be the sum of the binary cross-entropies on each of the $19$ target ICD codes. The hybrid model is trained using the ADAM optimizer \cite{adam}, with a learning rate of $10^{-3}$ and a batch size of $64$.

Of the $46,520$ patients in the MIMIC III dataset, we kept $10,000$ patients with at least one stay with 5 distinct ICD codes for validation and test, the remaining patients were attributed to the training set. As a result, the $58\,976$ patient stays of the MIMIC-III dataset is split into a training set of $44\,147$ stays, a validation set of $7\,472$ stays and a test set of $7\,357$ stays. The training set has $1.6M$ text documents. A summary is given below:
\begin{table}[h!]
  \begin{center}
\begin{tabular}{ |c|c|c|c|c| } 
 \hline
  &  total& train& validation& test \\ 
 \hline
 Patients &  46520&36520&5000&5000 \\ 
 \hline
 Stays & 58976& 44147& 7472& 7357\\
 \hline
 Documents&2083180&1661586&219751&201843\\
 \hline
\end{tabular}
  \end{center}
  \caption{Summary of train, validation and test sets.}
\end{table}
We stop the training when the loss on the training set stops decaying, which typically happens after a few hours of training on a GPU. Upon completion of the training, we extract an embedding for each stay by concatenating the $256$-dimensional the hidden layer of the MLP with the $192$-dimensional representation obtained with the CNN. We therefore obtain a $448$-dimensional representation of a stay, summarizing both the textual data contained in the medical reports and the structured data associated to the stay.

\section{Results}
\label{results_section}

In the following, we present our results both on the multi-label ICD code classification task and on the relevance of the learned embedding.


\subsection{ICD codes prediction}

As a first analysis, we show the performance of our algorithm for the prediction of the presence or absence of ICD codes at the lowest resolution level (i.e. for the $19$ ICD chapters). Results are provided in Table \ref{tab_pre_rec_f1}. In order to get a sense of the performance of our algorithm, we compare it a baseline random forest (multi-output) classifier trained on the raw structured features. More precisely, we train the random forest classifier on the training + validation set ($51,619$ visits). For all models, we compute precision, recall and $F_1$ score on the test set. The last column of Table \ref{tab_pre_rec_f1} gives the fraction of presence of each code on the test set (which are the roughly the same as in the whole dataset). We note a high imbalance in the labels classes, with ICD chapters presence ranging from $0.3\%$ for Complications of Pregnancy to $71.8\%$ for Diseases of the Circulatory System.

\begin{table}[h!]
  \centering
  \begin{tabular}{l|ccc|ccc|ccc|c}
\toprule
    {} & \multicolumn{3}{c}{Precision}& \multicolumn{3}{c}{Recall}& \multicolumn{3}{c}{$F_1$}\\
{} &    rf &  deep & emb+rf &    rf &  deep & emb+rf &    rf &  deep & emb+rf & Presence  \\
\midrule
Diseases Of The Circulato... & 0.841 & 0.994 &  0.963 & 0.998 & 0.999 &  0.997 & 0.891 & 0.997 &  0.980 &      0.718 \\
Endocrine, Nutritional An... & 0.727 & 0.746 &  0.730 & 0.995 & 0.930 &  0.952 & 0.813 & 0.828 &  0.826 &      0.595 \\
Supplementary Classificat... & 0.671 & 0.742 &  0.666 & 0.982 & 0.748 &  0.767 & 0.741 & 0.745 &  0.713 &      0.572 \\
Diseases Of The Respirato... & 0.834 & 0.996 &  0.974 & 0.609 & 0.997 &  0.991 & 0.681 & 0.996 &  0.982 &      0.418 \\
Injury And Poisoning         & 0.956 & 0.702 &  0.647 & 0.406 & 0.611 &  0.419 & 0.515 & 0.653 &  0.508 &      0.387 \\
Diseases Of The Genitouri... & 0.816 & 0.995 &  0.975 & 0.493 & 0.994 &  0.981 & 0.582 & 0.995 &  0.978 &      0.366 \\
Diseases Of The Digestive... & 0.939 & 0.992 &  0.980 & 0.397 & 0.995 &  0.951 & 0.516 & 0.994 &  0.965 &      0.354 \\
Symptoms, Signs, And Ill-... & 0.778 & 0.614 &  0.568 & 0.329 & 0.497 &  0.326 & 0.442 & 0.549 &  0.414 &      0.341 \\
Diseases Of The Blood And... & 0.762 & 0.993 &  0.967 & 0.304 & 0.995 &  0.908 & 0.420 & 0.994 &  0.936 &       0.325\\
Mental Disorders             & 0.774 & 0.559 &  0.531 & 0.126 & 0.271 &  0.066 & 0.215 & 0.365 &  0.117 &       0.279\\
Supplementary Classificat... & 0.965 & 0.739 &  0.701 & 0.144 & 0.430 &  0.150 & 0.237 & 0.544 &  0.247 &      0.278 \\
Diseases Of The Nervous S... & 0.969 & 0.996 &  0.995 & 0.141 & 0.990 &  0.820 & 0.231 & 0.993 &  0.899 &      0.263 \\
Infectious And Parasitic ... & 0.996 & 0.711 &  0.636 & 0.291 & 0.501 &  0.241 & 0.417 & 0.588 &  0.350 &      0.245 \\
Diseases Of The Musculosk... & 1.000 & 0.987 &  0.988 & 0.002 & 0.973 &  0.338 & 0.005 & 0.980 &  0.504 &      0.168 \\
Neoplasms                    & 1.000 & 0.680 &  1.000 & 0.015 & 0.168 &  0.001 & 0.030 & 0.269 &  0.002 &      0.151 \\
Diseases Of The Skin And ... & 1.000 & 0.455 &  0.000 & 0.001 & 0.040 &  0.000 & 0.003 & 0.074 &  0.000 &      0.101 \\
Certain Conditions Origin... & 1.000 & 0.999 &  1.000 & 0.504 & 1.000 &  0.958 & 0.624 & 0.999 &  0.979 &     0.093 \\
Congenital Anomalies         & 0.895 & 0.992 &  1.000 & 0.082 & 0.960 &  0.398 & 0.148 & 0.976 &  0.570 &     0.051 \\
Complications Of Pregnanc... & 0.000 & 1.000 &  0.000 & 0.000 & 0.792 &  0.000 & 0.000 & 0.884 &  0.000 &     0.003 \\
\midrule
Total average                & 0.737 & 0.860 &  0.841 & 0.498 & 0.784 &  0.683 & 0.595 & 0.820 &  0.754 &             - \\
\bottomrule 
\end{tabular}

  \caption{Precision, Recall, F1 scores for a random forest classifier (rf) trained on the raw data, our deep neural network (deep) and a random forest classifier trained on the embeddings (emb+rf).}
  \label{tab_pre_rec_f1}
\end{table}

                We see that our algorithm has a better $F_1$ score for all chapters. The global precision and recall of our deep neural network is also better than those of the random forest classifier.

                In order to test the quality of our embeddings, we also train the random forest classifier on the embeddings and the results are given in the column `'emb+rf`' in Table \ref{tab_pre_rec_f1}. As expected, the performances are lower than the predictions made by our neural network. However, we see that on average the performances of the classifier are higher when trained on the embeddings than on the raw data. This shows that our embeddings which are now of much lower dimension than the raw data still contain the relevant medical information (at least to make good ICD codes predictions).

               \subsection{Medical semantic information encoding}
                
               We also assess the quality of our embeddings by looking at how medical concepts are encoded in the embedding space. In the following, we concentrate on antibiotic resistance and shock.
Our network produces a low-dimensional embedding from which ICD codes are predicted. As a result, the network compresses the EHR data in an efficient way. Indeed if a medical concept is very important to predict ICD codes, we should be able to recover it from our embeddings. We show that this is indeed true and that antibiotic resistance (resp. shock) corresponds to a particular dimension of our embeddings.

\begin{figure}[h!]
\centering
\includegraphics[scale=0.4]{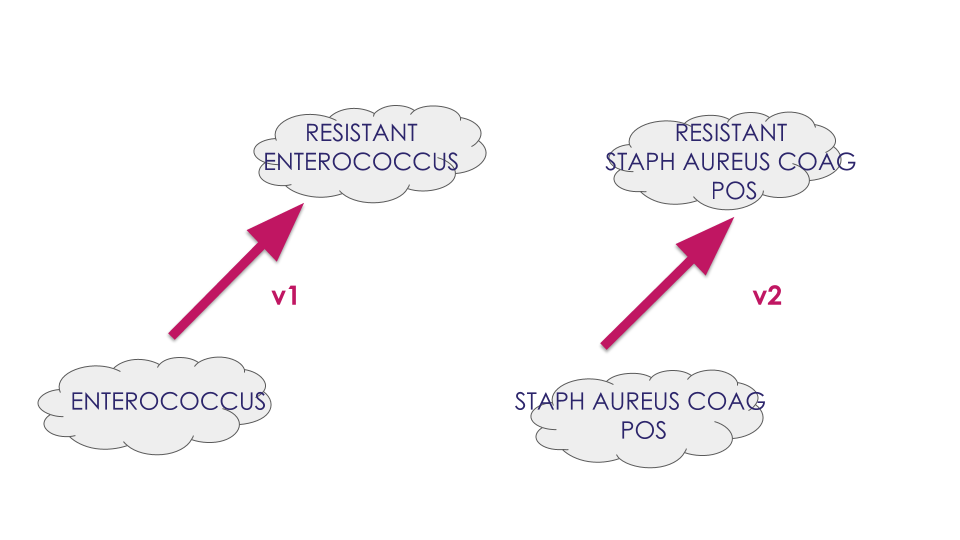}
\caption{Encoding of antibiotic resistance in our embeddings}
\label{translation}
\end{figure}
\begin{figure}[h!]
\centering
\includegraphics[scale=0.5]{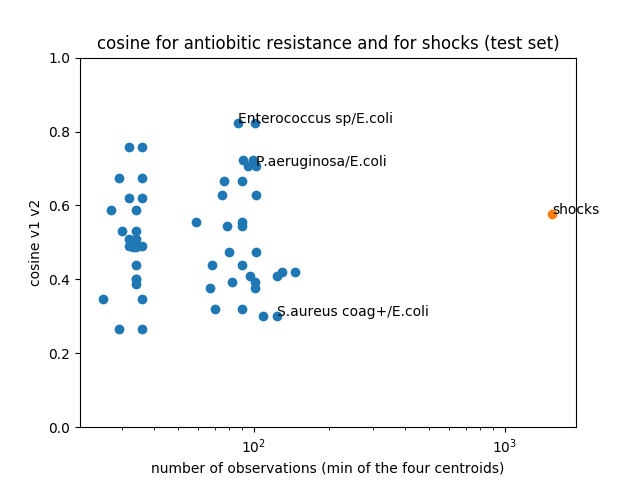}
\caption{Encoding of medical concepts in our embeddings: antibiotic resistance and shock}
\label{cos}
\end{figure}

In order to determine this particular dimension, we proceed as follows. Let us first consider two bacterias: Enterococcus sp. and Staph Aureus Coag.
To evaluate antibiotic resistance we then define four groups as pictured on figure \ref{translation}:
\begin{itemize}
 \item (a) stays with sensitive Enterococcus sp. microbiology result
 \item (b) stays with resistant Enterococcus sp. microbiology result
 \item (c) stays with sensitive Staph Aureus Coag. microbiology result
 \item (d) stays with resistant Staph Aureus Coag. microbiology result
\end{itemize}
The semantic relationship between groups (a) and (b) (resp. (c) and (d)) is antibiotic resistance.
We then compute the average vector $v1$ between groups (a) and (b) in our embedding and the average vector $v2$ between groups (c) and (d), as illustrated on Figure \ref{translation}.
Since each group has a different cardinality (with the smallest group (b) having $87$ stays), we compute the centroid of each group to properly define the vectors $v1$ and $v2$. Then, we compute the cosine of the angle $\langle v1, v2\rangle$ which in this case is $0.394$. As a comparison, two random vectors in our embedding space would have a cosine of mean $0$ and variance $\approx0.05$. We see that $v1$ and $v2$ are therefore significantly aligned. Indeed, we present in Figure \ref{cos} the corresponding cosines computed for each of the $54$ pairs of bacteria with at least $25$ stays. We see that none of the cosine is negative and indeed alignment is very significant with cosine values ranging between $0.265$ and $0.824$.


We carried out the same experiment for shocks, defining the four following groups:
\begin{itemize}
  \item (a) sepsis (as defined with angus criterium)
  \item (b) septic shock (angus criterium and vasopressor)
  \item (c) acute myocardial infarction (ICD9 code $410$)
  \item (d) cardiogenic shock ($410$ and vasopressor)
\end{itemize}
The semantic relationship between (a) and (b) is similar to that between (c) and (d), and relates to the notion of shock. We compute the four centroids of stays in the embedding space for each group. We then use the cosine between the vectors (a), (b) and (c), (d) as a similarity measure. The less populated of the four groups is (d) and has $1,548$ stays. We find a cosine value of $0.577$, implying a significantly better alignment than two random vectors.

\section{Discussion}

One strong advantage of our method is that the representation is learnt without any human intervention where traditional approaches require a clinician expert to determine patterns to look for in the tens of thousands of features and free text included in EHRs as well as computer science expertise to implement them. Despite the notorious difficulty to interpret embeddings, we managed to experimentally prove that clinical concepts such as antiobiotic resistance or shocks are encoded in the resulting vector space. The joint learning allows to leverage information contained both in free text and structured data. This is an important aspect as most of the clinical information can be included only in free text: for example, more than $85\%$ of the auto-immune comorbidities in \cite{escudiebmcmidm2017} were not present in the structured data. Other published deep learning works focused either on structured data only\cite{med2vec} \cite{deeppatient} or free text data only \cite{txt1} \cite{txt2}.

Whereas previous works show that the underlying neural networks can be trained in a self-supervised way -- see the stacked autoencoders in deep patient\cite{deeppatient} -- we chose here a supervised multiclass classification of ICD9 diagnosis chapters, motivated by the goal of learning a general embedding encompassing the whole spectrum of clinical semantics. This is a hard classification task for many reasons, especially because a stay can have multiple classes, but also because of the high class imbalance even at the highest level of the ICD9 classification. At this specific level -- corresponding to 19 broad classes -- our model however manages to organize finer concepts in its internal representation.

The information on antiobiotic resistance or the broader notion of shocks, whether septic or cardiogenic, although present in the input features, was indeed correctly reorganized as directions in the embedding space, as the colinearity measurement shows. This evaluation can be seen as an equivalent of word analogies for embeddings learnt on raw text\cite{word2vec} (Paris - France and Italy - Rome) and are seen as an indicative quality of the embedding.

The size of the MIMIC III database is a relative limitation to deep learning approaches where larger datasets typically allow for training larger neural networks to train and yield higher perfomance. This dataset has approximately $58,000$ stays and 2 millions documents, when a typical university hospital clinical data warehouse would store a few millions stays\cite{jannotijmi2017}. 
Deep learning can still be sucessfully used on MIMIC III \cite{deepmimic}. In this study, some design choices are driven by these limitations. We limit the text input to a vector of $2,000$ words, we aggregate data to one timestamp per stay, losing the temporal dimension, we use shallower neural network architectures, and we reduced the number of structured features from $40,000$ to $8,000$. The feature selection is automated and complies with our goal that no expert intervene on the feature learning step. It is done by simple chi-square test and is possible thanks to the choice of a supervised classification tasks as auxillary learning tasks.

This work describes an accessible method to produce embeddings on clinical data typically found in clinical data warehouses and advances the medical interpretation of the internal representation yielded by neural networks. It is quite computationally expensive like every deep learning method, but the reuse of the embedding in other tasks justifies the initial cost. Indeed, we show that the embedding improves the performance of less computationally intensive algorithms, such as random forest, on the ICD coding prediction task. Our exploration of the embedding structure also motivates its application to similarity-based patient retrieval. Precomputing the embedding for a large clinical data warehouse could then be useful for other classification tasks or cohort selection. Moreover, our embedding benefits from encapsulating variables dependencies learnt on a large set of stays, which could prove useful to develop an algorithm predicting a specific condition, typically on a smaller dataset.

\section{Conclusion}

We presented a neural network architecture that successfully takes raw EHR data, both structured and free text as input and automatically builds a stay representation as a buyproduct of learning, to predict the ICD9 diagnosis chapters assigned to each stay. We demonstrate that this embedding conserves a general medical semantic representation of the initial data. The embedding also improves the performance of a random classifier on a prediction tasks compared to the raw data on the MIMIC III dataset.

Our approach is general and flexible and opens the way to a lof of variations. More information could be incorporated into the learning task, and recurrent neural networks will be used to include time dependency. Moreover, different kind of informations can be withheld depending on the ultimate goal in order to modify the supervised learning task, which will lead to the study of the resulting embeddings.


\section*{Acknowledgements}
Most of this work was done during the Datathon DAT-ICU organized by AP-HP in Paris on January 20-21 by the team DLDIY. In addition to the authors, Fajwel Fogel (Sancare), Mina He (BNP Paribas) and G\'erard Weisbuch (ENS) contributed to the project and we thank them for their work. The team was one of the two winning projects\footnote{\url{https://www.aphp.fr/contenu/datathon-dat-icu-intensive-care-unit-4-projets-innovants-selectionnes-lissue-de-48h-danalyse}}

\makeatletter
\renewcommand{\@biblabel}[1]{\hfill #1.}
\makeatother

\bibliographystyle{unsrt}

\end{document}